# Active control of ion transport within a nanofluidic system

Sinwook Park and Gilad Yossifon*

*Faculty of Mechanical Engineering, Micro- and Nanofluidics Laboratory, Technion – Israel Institute of Technology, Technion City 3200000, Israel*

\* Corresponding author: yossifon@technion.ac.il

## Abstract

The ability to induce regions of high and low ionic concentration adjacent to a permeselective membrane or nanochannel subject to an externally applied electric field (a phenomenon termed concentration-polarization) has been used for a broad spectrum of applications ranging from on-chip desalination, bacteria filtration to biomolecule preconcentration. But these applications have been limited by the ability to control the length of the diffusion layer that is commonly indirectly prescribed by the fixed geometric and surface properties of the nanofluidic system. Here, we demonstrate that the depletion layer can be dynamically varied by inducing controlled electrothermal flow driven by the interaction of temperature gradients with the applied electric field. To this end, a series of microscale heaters, which can be individually activated on demand are embedded at the bottom of the microchannel and the relationship between their activation and ionic concentration is characterized. Such spatio-temporal control of the diffusion layer can be used to enhance on-chip electro-dialysis by producing shorter depletion layers, to dynamically reduce the microchannel resistance relative to that of the nanochannel for nanochannel based (bio)sensing, to generate current rectification reminiscent of a diode like behavior and control the location of the preconcentrated plug of analytes or the interface of brine and desalted streams.

## Introduction

Passage of an electric current through a permselective membrane results in regions of depleted and enriched ionic concentrations at the opposite sides of the membrane interfacing an electrolyte solution, a phenomenon termed concentration polarization (CP). This phenomenon has been the focus of intensive research in the last decade, particularly regarding its relation to microfluidic applications[1,2], e.g. on-chip desalination where the CP layer is used to separate between brine and desalted streams[3], or enhanced immunoassay sensing by preconcentration of analytes at the edge of the depletion layer, resulting from a field-focusing-gradient effect[4].

For low applied voltages, the current through the membrane is Ohmic (i.e. linearly increasing with applied voltage) until the diffusion-limited current saturates when both ion concentrations are completely depleted at the surface. The limiting current density scales as the inverse of the diffusion layer length and for an ideal, 1D permselective membrane with negligible convection, the exact relation is $i = 2zFDc_\infty/L$ [5], where $z$ is the valency, $F$ is the Faraday constant, $D$ is the diffusion coefficient, $c_\infty$ is the bulk ionic concentration and $L$ is the diffusion layer length. Unless there is some kind of convective stirring in the system, the diffusion length spans the entire distance, from the membrane interface to either the electrode or the reservoir. Since propagation of the depletion layer results in increased system resistance, its chronopotentiometric response shows a monotonic increase in the voltage. Saturation of the voltage occurs when the diffusion layer reaches its finite length[6].

In applications requiring intense ion transport, e.g. electrodialysis, it is desirable to shorten the diffusion length. It has previously been shown that saturation in macroscale systems can be induced much faster by natural[7] or forced [8–12] convection that suppresses the diffusive growth, resulting in a smaller diffusion layer. On the microscale, scaling arguments[13] suggest that natural convection becomes less important with decreasing characteristic dimensions, while use of forced convection is practically less favorable due to the increased hydrodynamic resistance and the need to combine an external pump. In a heterogeneous system (i.e., membranes with a partly conducting surface area[14] or fabricated micro-nanochannel systems[15,16]), the diffusion layer length may be controlled by electroosmotic flow (EOF) of the second kind,[17] that induces electro-convective stirring, as seen in both fabricated nanochannels[16,18] and heterogeneous membranes[19]. In both of these heterogeneous systems, a significant tangential component of the electric field exists along the membrane interface, and drives the extended space-charge (ESC) that is induced by the normal component of the field[20]. In contrast, in the case of homogenous (i.e. pseudo 1D) permselective systems, which have a negligible tangential field component, in the absence of external stirring, electro-convection only emerges after surpassing a certain voltage threshold in the form of an electro-convective instability[21,22]. This electrokinetic instability evolves into a stationary interfacial vortex array that is found to arrest the self-similar diffusive front growth[23,24], which in turn, specifies the overlimiting current (OLC). Competition between these two mechanisms, EOF of the 2$^{nd}$ kind and electro-convective instability, has also been recently shown[15,25] to control the diffusion layer length.

In all of the above-mentioned cases, the length of the diffusion layer is indirectly prescribed by the complex competition between several mechanisms which are primarily dictated by the system parameters (e.g., geometry, concentration, surface charge) and applied voltage. In contrast, we recently demonstrated a novel means of directly controlling the diffusion layer length, independently of the dominating OLC mechanism and system parameters, using alternating-current electroosmosis (ACEO)[26]. This was realized by embedding an array of non-insulated and individually addressable electrodes at the bottom of the microchannel. Two modes of induced-charge electrokinetic (ICEK) effects can occur, ACEO[26] or induced-charge electroosmosis (ICEO)[27], depending on whether the electrodes are powered (i.e., their potential is externally controlled) or floating, respectively. Thus, ICEO is considered a passive and ACEO an active mode of controlling the CP layer. Both arise from the action of the tangential component of an applied electric field on the electric double layer (EDL) induced by the same electric field over a polarizable surface. In both cases, vortices formed over the electrodes effectively stir the fluid, arresting the diffusive growth of the depletion region and controlling the limiting and overlimiting currents. However, the main deficiency of the ICEK mechanism is its suppression with increasing solution conductivity[28]. Furthermore, a possible complication with such non-insulated electrodes is Faradaic (redox) reactions that occur when a sufficiently high driving voltage is applied, causing the electrodes to act as bipolar electrode (BPE)[29,30].

Herein, we propose a novel mode of active control of the diffusion layer length that resolves the above-mentioned deficiencies, by driving fluid flow through electrothermal (ET) forces. These result from the interaction between the electric field and temperature gradients (i.e. permittivity and conductivity gradients)[13,31]. This is realized by embedding an array of thin film microheaters within the microchannel interfacing a permselective medium (Fig.1) to generate a non-uniform temperature field. In order to actively drive electro-convection by the electrothermal forces only, the electrodes are electrically insulated from the electrolyte via a thick dielectric coating. The dominance of ET flow at high conductivities and frequencies over ACEO has been demonstrated[31–34]. In addition, the insulation layer suppresses unwanted Faradaic reactions. The effect on diffusion length was estimated by visualizing the dynamics of the CP layer along with colloid dynamics and by measuring the chronopotentiometric response. The experimental results were qualitatively validated by numerical simulations with a fully coupled two-dimensional (2D) time-dependent model. Finally, a dynamic and periodic control of the depletion layer was demonstrated by turning on/off selected heaters on demand. To the best of our knowledge, harnessing ET for control of the CP layer dynamics has not been addressed in the past. Such novel spatio-temporal control of the depletion layer is expected to be of utility in on-chip electrodialysis and CP-based desalination applications.

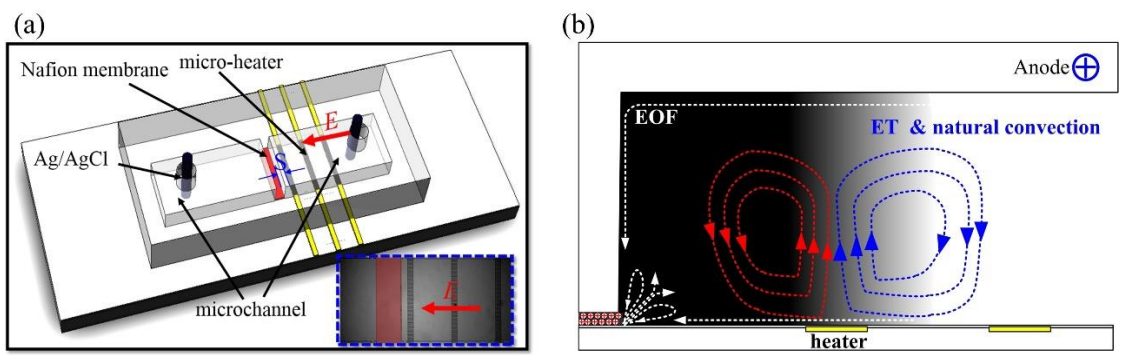

**Fig.1: Working principle of the ET-controlled diffusion layer within a microchannel-membrane device**. (a) A schematic of the microchannel–Nafion membrane device with an embedded array of microheaters for ET control. The spacing between the first microheater and the membrane is indicated as *S*. The inset indicates a microscopic image (top view) of a fabricated device. (b) A schematic illustration of the suppression of diffusion layer propagation by the induced ET flow and natural convection.

## Results

### *Characterization of the temperature field and electrothermal flow*

In order to measure the heater temperature as a function of the supplied power, we first characterized indirectly (Fig.2a-c), using a temperature-sensitive dye within the electrolyte, and then directly (Fig.2d), using an infra-red (IR) camera without an electrolyte. Recordings indicate linear dependency of the heater temperature on the applied power. In order to isolate the ET-generated vortices from linear EOF, we first evaluated the performance of a membrane-less microfluidic chip subjected to an AC field. The velocity profiles, extracted by particle tracking at various z-plane, are shown in Fig.2e. Comparison of the measured velocities to those obtained from natural convection ($E_0 \approx 0$) suggested that while the latter contribution is not negligible[13] under the studied conditions, with increasing electric fields, the ET dominated natural convection due to its quadratic dependency on the electric field (Fig.2f).

To describe the induced ET flow, the ET force under AC electric field conditions can be expressed as[31]

$$f_{ET} = \frac{1}{2}\varepsilon\left[\frac{1}{2}(\alpha-\beta)(\nabla T \cdot E)E - \frac{1}{2}\alpha|E|^2\nabla T\right], \tag{1}$$

where $\varepsilon = \varepsilon_r\varepsilon_0$ is the fluid permittivity ($\varepsilon_r$ is the dielectric constant while $\varepsilon_0$ is the vacuum permittivity), $\sigma$, $E$, $T$ stand for the permittivity, conductivity of the solution, electric field and temperature, respectively, and $\alpha = (1/\varepsilon)(\partial\varepsilon/\partial T) \approx -0.4\%\,°C^{-1}$, $\beta = (1/\sigma)(\partial\sigma/\partial T) \approx 2\%\,°C^{-1}$. It is clear that at DC field conditions, the Coulombic contribution (first term) to the ET force is dominant relative to the dielectric contribution. A scaling analysis for the ET velocity, based on the Stokes equation consisting of the ET force (eq.(1)), yields

$$u_{ET} = 0.25\varepsilon(\alpha-\beta)E_0^2(\nabla T)d^2\eta^{-1}, \tag{2}$$

where $d$ is the height of the microchannel and $\eta$ is the dynamic viscosity of the fluid. Applying eq.(2) to the experimentally measured ET velocities (Fig.2f) yields a fitted temperature gradient of (8 Kmm$^{-1}$, similar to the experimentally measured values (6.1-12.3 Kmm$^{-1}$) obtained by averaging the temperatures within a region of 100-250μm from the edge of the heater operating at 62mW. The fact that the quadratic scaling, $E_0^2$, for the ET velocity, $u_{ET}$ (obtained by subtracting the natural convection velocity, $u_{NC}$, from that measured total velocity) yielded the best fit, compared to $E_0$ or $E_0^4$, supports the claimed ET-induced flow mechanism that is due to the combined action of the external electric field and the temperature gradients generated by the heater. The respective range of Peclet number $0.8 \leq Pe_{ET} = u_{ET}d/D \leq 7$ (wherein $D$ is the diffusion coefficient), indicates that the ET-induced vortices may be strong enough to mix the solution nearby and in turn, modify and possibly suppress the diffusion layer growth.

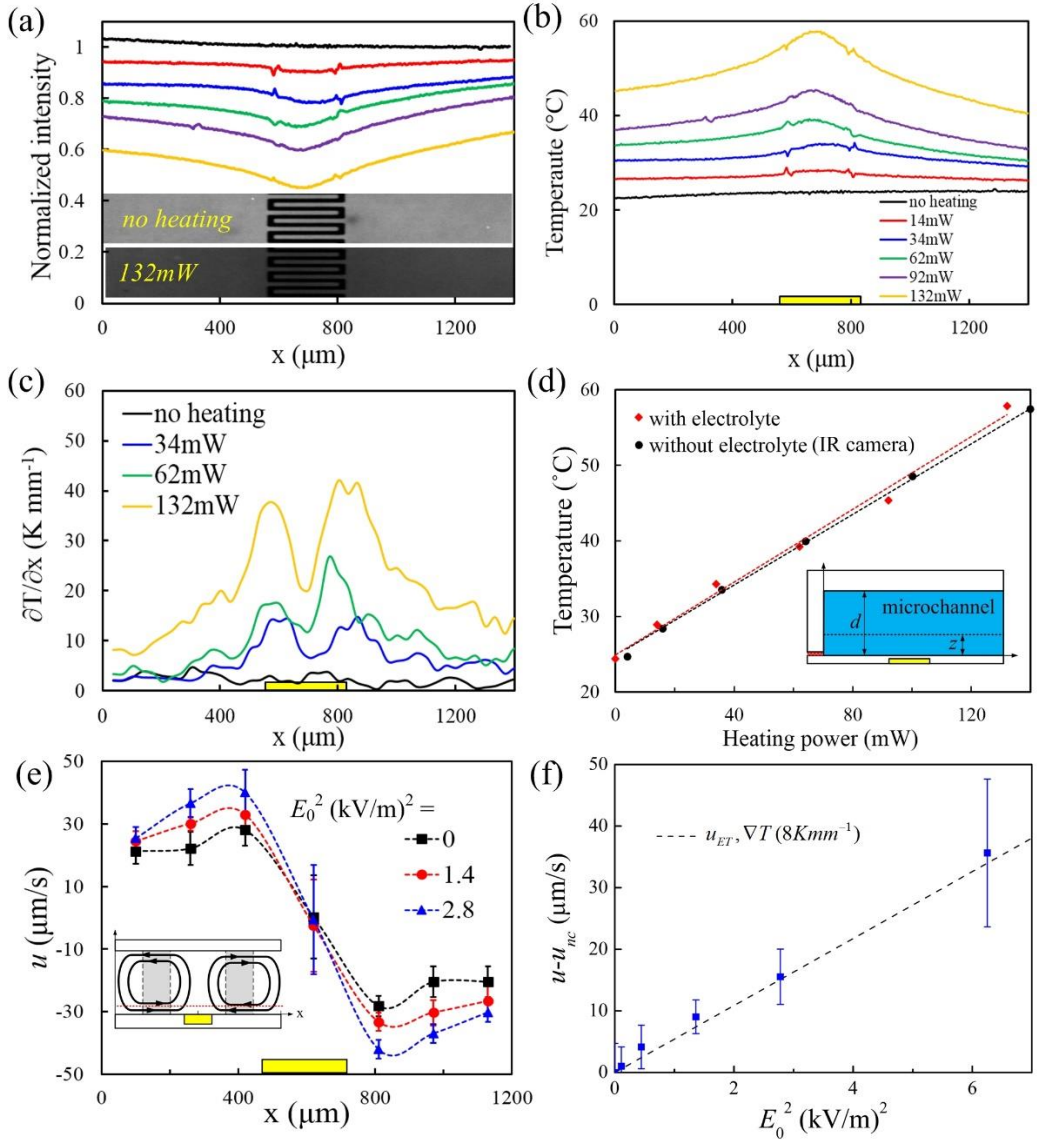

**Fig.2: Characterization of the temperature distribution induced by the embedded heaters and the resulting natural convection and ET flow for a simplified device consisting of a microchannel without a membrane.** (a) Microscopic images and normalized intensity profiles of the rhodamine B fluorescent dye at $\tilde{t}$ = 400s, for varying heater powers within a microchannel ($d$ = 1000 $\mu$m, $z_1$ = 100 $\mu$m). (b) The corresponding correlated temperature and (c) temperature gradient distributions. (d) The corresponding maximum temperature dependency on the heater power, as extracted via IR camera (without electrolyte) and rhodamine B fluorescent dye (with electrolyte from part b). The inset presents a scheme of measured depths of focal plane $z$, and the microchannel ($d$). (e) The measured velocity

component along the x-axis, $u$, within a microchannel ($d = 400$ μm, $z_1 = 100$ μm) as a function of various applied AC field amplitudes at a frequency of 1 kHz and a fixed heating power of 60 mW. The inset indicates the focal plane ($z_1 = 100$ μm) and the regions (grey rectangles - located 100-250 μm from the edge of the heater) at which the average velocities were measured. (f) The average measured velocity as a function of $E_0^2$ (blue markers) and scaling analysis of the ET velocities (eq.(2)). Herein, $u_{NC}$ is the measured average velocity of the natural convection (21 ± 4.7 μm s$^{-1}$). Quadratic scaling, $E_0^2$, yielded the best fit, compared to $E_0$ or $E_0^4$.

*The effect of electrothermal and natural convection flow on the concentration-polarization layer*

To understand the effect of ET and natural convection (NC) flow on the CP layer, we began by characterizing the microchannel-membrane system without heaters. As seen in Fig.S.3a-b, the limiting currents were proportional to the microchannel depth, whereas the chronopotentiometric response (Figs.S.3c-e) of the various systems exhibited the expected inflection point that corresponds to the Sand time[8] and scales with the inverse of the current density squared (Fig.S.3f). Suppression of depletion layer growth due to ET- and NC-induced flow was clearly enhanced with increasing microchannel depth and increasing applied heater power (Fig.3, Fig.S4). This was most pronounced with a 1-mm-deep microchannel (Fig.3h,i), whereas for shallower channels, although the depletion layer was affected by the ET force and NC flow, its growth was not arrested by the convective stirring. As clearly seen in the 1-mm-deep channel, besides arresting the CP growth, the depleted layer concentration increased with increasing heater power due to the enhanced convective stirring. These effects were also reflected in the chronopotentiometric response of the system. In particular, the continuous propagation of the depletion layer with the shallower microchannels, even at the highest heating powers, resulted in a corresponding continuous increase of the microchannel resistance, whereas at the 1 mm microchannel depth and sufficiently high (60 mW) heating power, the V-t response saturated very quickly (Fig.3g). Increasing the applied currents resulted in increased depletion (Fig.S5-7), whereas for shallower microchannels (i.e. 330μm), it overwhelmed the convective stirring and resulted in less effective suppression of depletion layer growth.

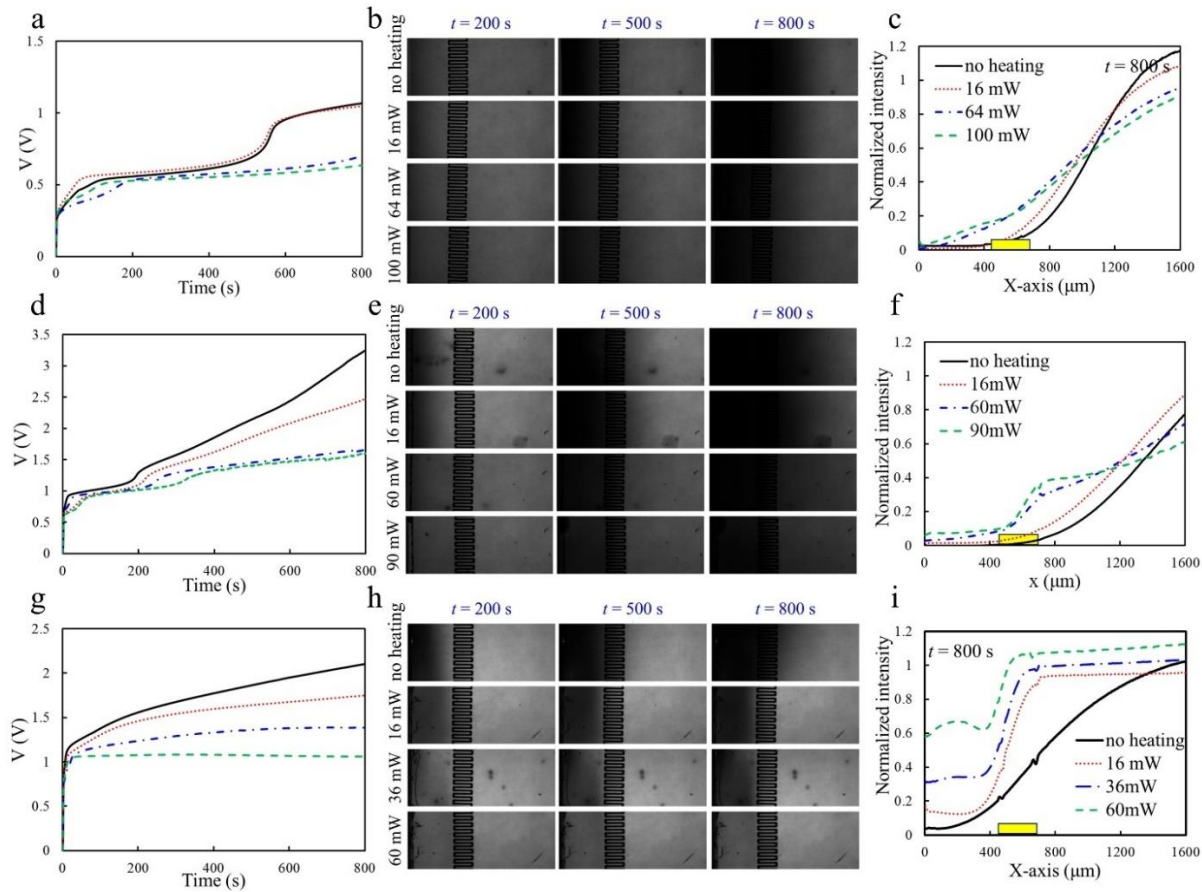

**Fig.3: The effect of ET- and NC-induced flow on depletion layer growth within membrane-microchannel systems with varying microchannel depths, *d*, for various applied heating powers**. (a, d, g) Chronopotentiometric (*V-t*) response at a fixed current regime ($1.5 \cdot I_{lim}$), which were 75, 375, and 720 nA for 110, 330, and 1000 $\mu$m-depth channels, respectively. (b, e, h) Time-lapse ($\tilde{t}$ = 200, 500, and 800s - see movies S1-S3) fluorescent images of the anodic side of the microchannel-membrane interface (*x*=0). (c, f, i) Corresponding normalized (by the bulk value) fluorescent intensity distribution at $\tilde{t}$ = 800 s. As clearly shown, with increasing channel depth, convective stirring further suppressed diffusive layer growth.

A qualitative understanding of the effect of the ET force and NC flow on the depletion region can be obtained by solving the Poisson-Nernst-Planck-Stokes (PNPS) equations along with the energy equation using a fully coupled, two-dimensional (2D) transient model (see in Supplementary for simulation details). The ET- and NC-induced flows were imposed in the Stokes equation as an electrothermal and buoyancy body force, respectively, and solved with either decoupled or combined ET and NC effects (Fig.4). When considering the ET flow, the depletion layer was only affected when passing over the heater (Fig.4d), thereby suppressing its growth. This results from the fact that the electric field is inversely proportional to the local concentration ($E \propto 1/c$), and hence, increases dramatically within the depletion layer. This can be seen nicely in Fig.S9, wherein the vortices associated with the heater increased as the depletion layer approached the heater. When combined with NC, an additive effect on depletion layer suppression was observed (Fig.4d), as both act in the same direction (i.e., the resulting vortices are of the same rotational direction, Fig.2e and Fig.4c). The numerical results of the temporal CP growth (Fig.4d) were in qualitative agreement to the experimental results of Fig.3, particularly that of the intermediate channel depth of 330μm (Fig. 3f), wherein suppression of the depletion layer occurred when it overlapped the heater.

To verify that the decreased system resistance, as seen in the V-t response of Fig.4e, was due to convective mixing rather than to the increased ionic diffusivity resulting from the elevated temperatures, the V-t response of the electro-diffusive temperature-dependent ionic transport in quiescent fluid was calculated and compared against the V-t response with ET flow (Fig.4f). Although there was a slight voltage drop with increased power in the former cases, the ET flow clearly dominated the voltage drop. To further verify this, experimental V-t responses under isothermal elevated temperatures (without operating the heater but rather using a hotplate) were recorded (Fig.4g). As expected, the uniform increase in the fluid temperature resulted in a voltage decrease due to the temperature-dependent diffusivity of ions, however the reduction was relatively small and cannot explain the much larger drop associated with ET forces.

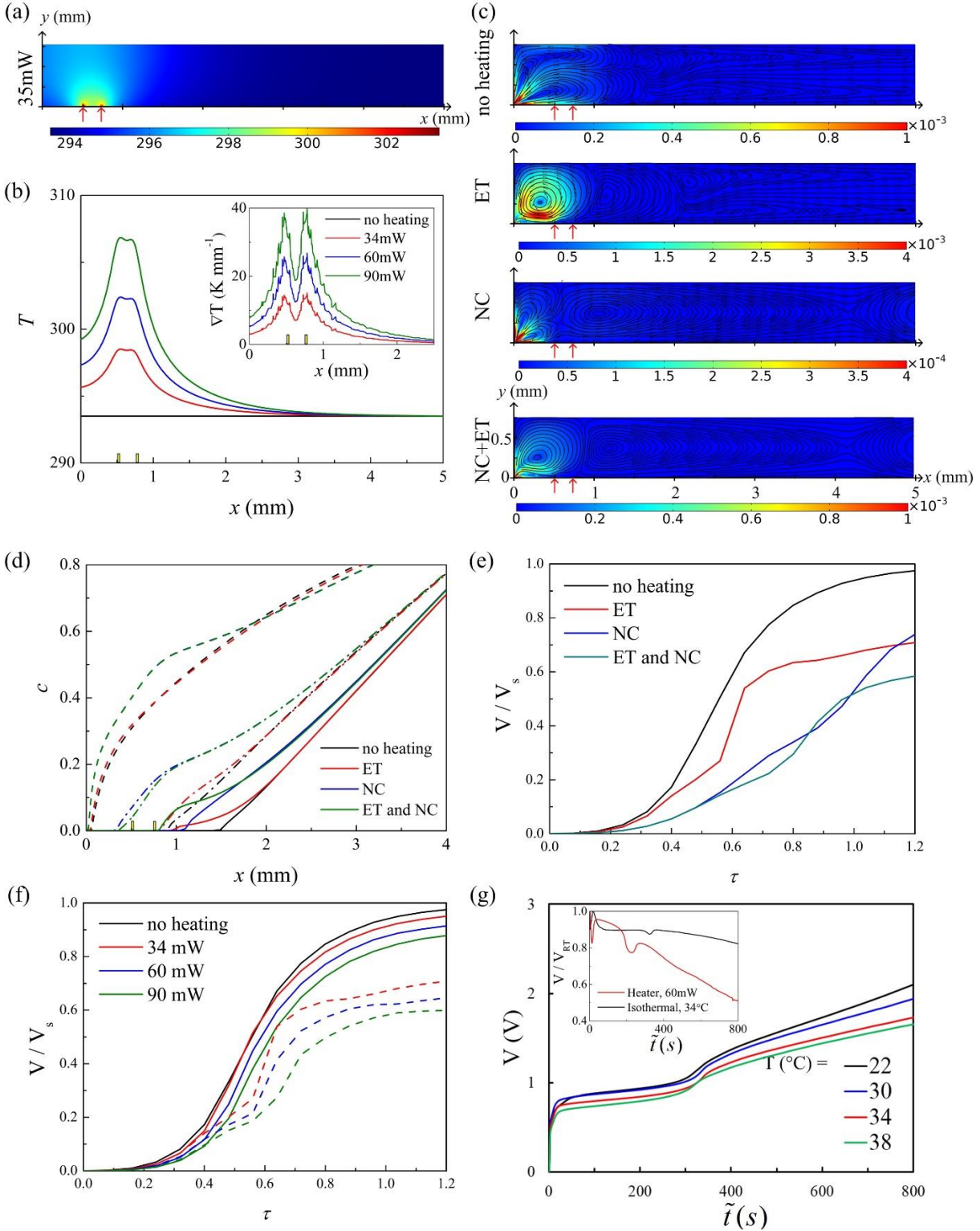

**Fig.4: Numerical simulations and examination of the effect of elevated temperature on the V-t response.** (a) Temperature field and (b) temperature distribution as functions of heating powers ($z_1 = 100$ $\mu$m) and their corresponding gradients (inset). (c) Velocity field for various combinations of convective

modes (EOF, NC, ET) at $\tau = (\tilde{t}/\tilde{t}_d) = 0.72$, $\tilde{t}_d = L^2/D$. The applied heating power was 34mW. Black lines and arrows indicate the flow streamlines and velocity vectors, respectively, while the two red arrows point to the locations of the 25 μm-wide heater lines ; (d) Time evolution of the depletion layer in terms of its salt concentration distribution, $c = (c_+ + c_-)/2c_0$ at $\tau = 0.24$ (dashed), 0.64 (dashed-dotted) and 1.2 (solid), for the various convective modes. (e) The corresponding normalized chronopotentiometric (V-t) responses by the steady-state voltage at no heating ($V_s$). In all cases, the applied current density was $i = 1.45\ i_{\text{lim, 1D}}$ (where $i_{\text{lim,1D}}$ stands for the limiting current density in the 1D case) and EOF was accounted for by using a zeta potential of $\zeta = -10\text{mV}$. For the parameters of diffusion coefficient of ion pairs, viscosity, permittivity and thermal conductivity of fluid, the temperature-dependent physical properties were considered. A microchannel depth of 750μm was used in the simulation. (f) Simulated V-t response for varying heating powers, with temperature dependent material properties but without convection (solid lines) and with ET and EOF (dashed lines). (g) Experimental V-t response under isothermal conditions at various elevated uniform temperatures. The inset depicts the voltages (normalized by the value obtained at isothermal 22 ºC without heater conditions) measured for the two cases of isothermal (34 ºC) without heater and for 60 mW heater within the membrane-microchannel system ($d$ = 330 μm).

*Dynamic control of the concentration-polarization layer*

To demonstrate the dynamic control of the CP layer via ET-induced flow, periodic step-wise heating power (60 mW) was applied at a fixed current (720 nA) for the 1mm depth microchannel system, which exhibited the most pronounced suppression of CP growth (Fig.3). This was measured both in terms of the chronopotentiometric response and the corresponding microscopic images of the depletion layer (Fig.5) for various time periods, with the no-heating case serving as control. It is clearly seen that when applying heat, the resistance of the system dropped almost to its Ohmic value and the concentration of the depletion layer was significantly increased due to the vigorous ET- and NC-induced mixing. Once the heating was turned off, the concentration within the depletion layer decreased again and, together with the continued propagation of the depletion layer away from the membrane interface, resulted in an increase in the resistance (i.e., voltage increase). With increasing frequency of the current signal, there was less time for the depletion layer to recover to the no-heating status, as clearly seen in the chronopotentiometric response.

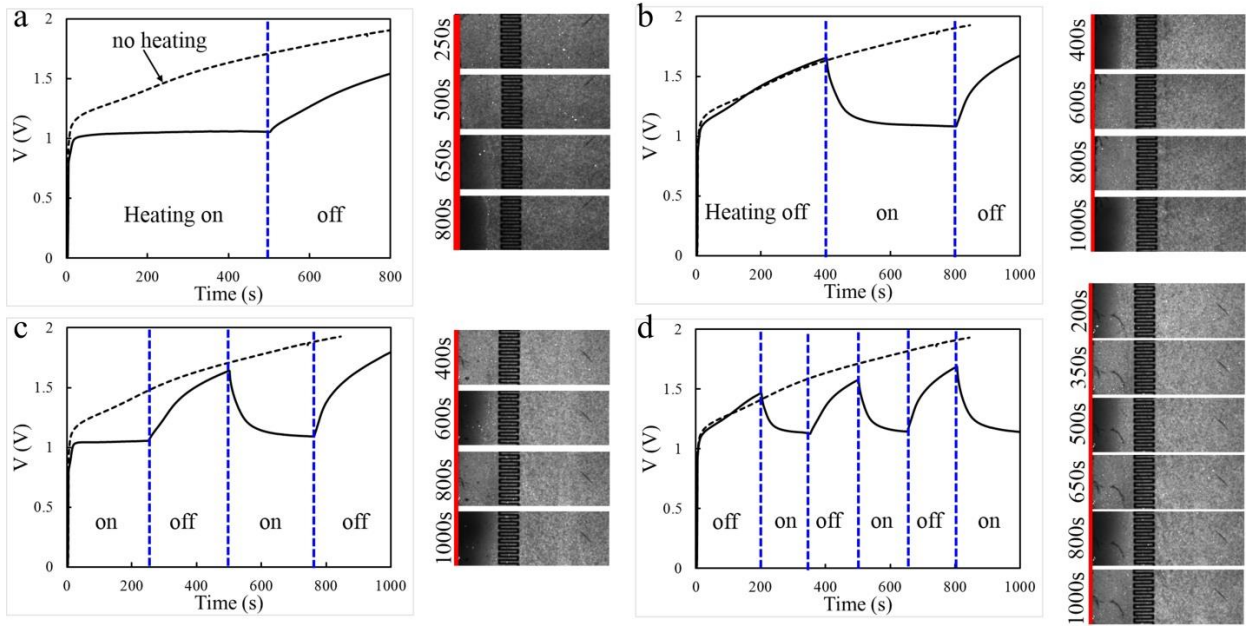

**Fig.5: Dynamic control of the CP layer.** V-t response along with its corresponding microscopic images for various frequencies of the periodic step-wise application of the heater (60 mW), at a constant applied current (720 nA) for CP within the 1mm-depth channel.

## Discussion

It has been demonstrated that ET-induced flow, resulting from a combination of an externally applied electric field and temperature gradients generated by local heaters embedded at the microchannel surface, combined with natural convection, can effectively modulate the response of a microchannel-membrane system via its mixing/stirring effect of the depletion layer. At sufficiently deep microchannel depths (~1 mm), these flows can fully arrest the growth of the depletion layer, along with increased ionic concentration within the depletion layer, due to increased mixing. This was also manifested by the resulting system resistance, which rapidly reached saturation, in contrast to the no-heating case, wherein the system resistance continuously grew in correspondence to depletion layer growth. In addition, it was shown that a periodic step-wise application of the heater, with the frequency as a control parameter, enabled control of the depletion layer and of the associated system resistance. At a sufficiently high frequency, the depletion layer is unable to recover to the no-heating status and hence limits the system resistance.

Such novel spatio-temporal control of the depletion layer is expected to be of utility in on-chip electrodialysis, where it is desirable to shorten the diffusion length for intense ion transport, and CP-based desalination where the division between the brine and desalted streams can be controlled via an array of heaters. It can also allow for control of the location of the preconcentrated plug of biomolecules at the outer edge of the depletion layer in immunoassay applications. Furthermore, asymmetric actuation and/or integration of the array of heaters at the opposite sides of the membrane (e.g., as in Fig.1, where one side is bare while the other has an array of heaters) leads to ionic current rectification (ICR) upon reversal of the external applied electric field, i.e. a nano-fluidic diode like behavior. In contrast to other mechanisms of rectification resulting from asymmetry of the permselective medium itself (e.g., asymmetric geometry as in nanopore[35] or funnel-shaped nanochannels[36,37]) the current mechanism is the direct consequence of the asymmetric depletion layer. However, in contrast to previous studies of the latter mechanism[38], wherein the current rectification value is determined by the system fixed geometry, the current ET-controlled CP enables dynamic variation of the rectification (see Fig. 5) as well as higher values of rectification due to enhanced control over the contrast between the CP responses at the opposite sides of the membrane.

## Methods

*Chip fabrication*

The device (Fig.1) consisted of a straight Nafion membrane (3mm in width, 1mm in length) flanked by two polydimethysiloxane (PDMS) microchannels (3mm in width, 8.25mm in length). In all experiments, the rectangular microchannels were of a constant lateral dimensions, while the depth was varied. The design was similar to previously studied microchannel-Nafion interface devices[39], with the embedded electrode array substituted with a serpentine heater array (20μm in width, 240μm in length, ~150Ω) at one side of the microchannel. The patterned micro-heater supplies an external heating source to generate a temperature gradient, which is independent to the applied electrode field for CP generation. The Faradaic reactions or other electrochemical reactions above the heater were fully suppressed by coating multi-stacked electrical insulating layers using silicon oxide/silicon nitride (0.5μm/1μm in thickness). The opposite microchannel was left "bare" as a control to study current-voltage (I-V) characterization. The detailed fabrication process is presented discussed in our previous publication[39].

*Experimental setup*

Two Ag/AgCl, 0.38mm-diameter electrodes (A-M systems) were inserted within each reservoir (1.5mm in diameter with its center located at the end of the channels, which were ~8mm from the membrane interface), and connected to Gamry reference 3000 for CP generation. Details of the chip wetting and cleaning prior to experiments are described elsewhere[40]. The electrolyte used in all experiments was KCl (~0.7 mM; $\sigma$ = 108 μS/cm, pH = 6.13). The current−voltage (I−V) curves were obtained from linear sweep voltammetry with a slow sweep-rate of 0.1V every 100s, from 0 to 3V. For ionic concentration profiles, 10 $\mu$M concentrations of pH-free Dylight molecules (Dylight 488, Thermo Scientific) were mixed in the KCl electrolyte. Their measured fluorescent intensity was further analyzed by normalizing the local fluorescent dye intensity by that away from the interface. The microheater, connected to a DC power supply (Agilent 3612A), was activated with various heating powers. The electrothermally induced velocities were tracked using 2 $\mu$m-fluorescent particles (Thermo Scientific) of 0.01% volumetric concentration. All experiments were recorded with a spinning disc confocal system (Yokogawa CSU-X1) connected to a camera (Andor iXon3). An infrared camera (Optris PI 450) was used to monitor the temperature of the bare heater without a bonded microchannel above. Additionally, the maximum temperature of heater were extracted from a linear approximation of the resistance of the heater versus temperature, $R(T) = R(T_0)(1 + \alpha(T - T_0))$, where $R(T)$ and $R(T_0)$ are the resistance values at temperatures $T$ and $T_0$, respectively, and $\alpha$ is the temperature coefficient of resistance[41]. The local temperature distributions above the heater, that is immersed in the microchannel filled with KCl electrolyte, were measured using a fluorescent dye (10 μM rhodamine B, Sigma)[41,42]. The microscope focal planes ($z$) were 35, 110 and 110 μm from the bottom of the microchannel which had a depth ($d$) of 110, 330 and 1000 μm, respectively (for more data, see Supporting materials Fig.S2). The slight decrease of the fluorescent intensity in Dylight solution upon application of various heating powers proved insignificant to the overall concentration profiles (Fig.S10).

## Acknowledgements

We acknowledge funding by the Israel Science Foundation (ISF), grant number 1938/16. Fabrication of the chip was made possible through the financial and technical support of the Russell Berrie Nanotechnology Institute and the Micro-nano Fabrication Unit. We also thank Dr. Alicia Boymelgreen for her useful comments.